\documentclass[smallextended]{svjour3}
\smartqed  
\usepackage[fleqn]{amsmath}
\usepackage{graphicx}
\usepackage{url}
\usepackage{graphicx}
\usepackage{enumerate}
\journalname{Celestial Mechanics and Dynamical Astronomy}
\begin{document}

\title{Mass of the Kuiper Belt}

\titlerunning {Mass of the Kuiper Belt}

\author{E. V. Pitjeva
\and N. P. Pitjev}

\authorrunning {E. V. Pitjeva, N. P. Pitjev}

\institute{E. Pitjeva \at
Institute of Applied Astronomy of Russian Academy of Sciences,
Kutuzov Quay~10, 191187 St.Petersburg, Russia \\
          \email{evp@iaaras.ru}
          \and
           N. Pitjev \at
St.Petersburg State University, Universitetski pr. 28, Petrodvoretz,
198504, Russia; Institute of Applied Astronomy of Russian Academy of Sciences,
Kutuzov Quay~10, 191187 St.Petersburg, Russia; \\
\email{ai@astro.spbu.ru}\\
}

\date{Received: 13 December 2017 / Accepted: 24 August 2018
\\
\\
The final publication ia available at Springer via\\
\url{http://doi.org/10.1007/s10569-018-9853-5}
}

\maketitle

\begin{abstract}

 The Kuiper belt includes tens of 
thousands of large bodies and millions of smaller objects. The main part of the 
belt objects is located in the annular zone between 39.4 au and 47.8 au from the 
Sun, the boundaries correspond to the average distances for orbital resonances 3:2 
and 2:1 with the motion of Neptune. One-dimensional, two-dimensional, and discrete 
rings to model the total gravitational attraction of numerous belt objects are 
considered. The discrete rotating model most correctly reflects the real 
interaction of bodies in the Solar system. The masses of the model rings were 
determined within EPM2017---the new version of ephemerides of planets and the 
Moon at IAA RAS---by fitting spacecraft ranging observations. The total mass of the 
Kuiper belt was calculated as the sum of the masses of the 31 largest trans-neptunian 
objects directly included in the simultaneous integration and the estimated mass 
of the model of the discrete ring of TNO. The total mass is 
$(1.97 \pm 0.30)\cdot 10^{-2} \ m_{\oplus}$. The gravitational influence of the 
Kuiper belt on Jupiter, Saturn, Uranus and Neptune exceeds at times the attraction 
of the hypothetical 9th planet  with a mass of $\sim 10 \ m_{\oplus}$ at the 
distances assumed for it. It is necessary to take into account the gravitational 
influence of the Kuiper belt when processing observations, and only then to 
investigate residual discrepancies to discover a possible influence of a distant 
large planet.


\end{abstract}

\section{\bf Introduction}
The attention of researchers of the Solar system to the Kuiper belt has constantly increased 
in recent decades. Interest is connected with the discovery of new trans-Neptunian objects 
(TNOs) and the refinement of their distribution over distances and sizes. 
The belt shows a diverse structure of 
orbits evoking questions about its formation, the composition of the primary 
material, the early history of the Solar system. A more reliable account of the influence of the belt as a whole on 
the planets will also allow us to decide how to search a large planet for an 
alleged distant (Batygin and Brown 2016; Fienga et al. 2016; Folkner et al. 2016; 
Millholland and Laughlin 2017). 

 The aim of the work is to obtain an estimate of the mass of the Kuiper
Belt and to improve and refine the used dynamic model of the Solar system to
derive exact ephemerides of the planets and more adequately take into account
the influence of a large number of distant bodies concentrated behind the orbit
of Neptune. It is necessary to take into account the gravitational attraction
of trans-Neptunian objects (TNOs), primarily to the planets Neptune, Uranus, Saturn,
closest to the Kuiper belt, due to the fact that the number
of objects turned out to be large and the total mass according to statistical and
dynamical estimates is much higher than the total mass of the asteroid belt. The
influence of the gravitational attraction of the Kuiper belt on observations
is shown and discussed in the new Section 6. Observations of the planets, mostly the
high-precision observations of Cassini spacecraft obiting the Saturn,
as well as the new version of the ephemerides of planets and the Moon -- EPM2017 at IAA
RAS were used for estimating the Kuiper belt mass.

\section{\bf Statistical estimates of the mass of the Kuiper Belt}
Kuiper Belt objects are ice bodies that include frozen methane, 
ammonia, water, carbon dioxide surrounding stony internal parts. Therefore, the 
average density of the Kuiper belt objects is, as a rule, about 1.5-2 $g/cm^3$.

  Kuiper belt objects (KBOs) are divided into three main dynamic classes 
by the characteristics of their orbits (Jewitt et al. 1998; Gladman 2002): 
classical objects, resonant objects and objects of the scattered disk.

Classic KBOs have near-circular orbits lying in the region of 40-50~au from 
the Sun and having relatively small eccentricities. These objects are not
strongly
affected by large planets, their orbits remain practically unchanged. 
They are the most numerous and constitute the main part of the population of the belt. 
The plane of the Kuiper belt defined by the classical objects with slopes $|i| < 5^o$
coincides with the invariant plane of the Solar system (Elliot et al. 2005). 
The main Kuiper belt is located between the orbital resonances 3 : 2 and 2 : 1 
with Neptune in the annular belt 39.4~au $ < A < $ 47.8~au (De~Sanctis et al. 2001). 
In the distribution there is a so-called Kuiper Cliff -- 
a sharp drop of the number of classical objects after 50 au. 
More precisely, the number of objects larger than 40~km falls sharply (Jewitt et 
al. 1998, Trujillo and Brown 2001a; Gladman et al. 2001; Allen et al. 2002) at a 
distance of 48 au. That is, the outer boundary for the belt of classical objects 
is very sharp. It is assumed that perhaps it corresponds to the edge of 
the  primordial protoplanetary cloud. There is compaction in the distribution 
("kernel") with a concentration of orbits having semiaxes about $a \sim 44$ au, 
eccentricities $e \sim 0.05$ and slopes $|i| < 5^o$ (Petit et al. 2011; Bannister 
et al., 2016). 

Resonant KBOs are primarily objects that are in 
orbital resonance with a mean motion with Neptune 3 : 2 (plutino, $a \sim 39.4$ 
au) and 2 : 1 ($a \sim 47.8 $ au), although there are a number of bodies with other 
resonant ratios of average motions.

The most part of scattered objects having orbits stretching far beyond 50~au, 
(from $a \sim 90$ au) and $e \sim 0.5-0.6 $ and more. The objects of the scattered 
disk are objects with large eccentricities and large slopes.
The formation of such a structure of the belt is explained by perturbations from 
the planets and, first of all, from Neptune (Levison and Morbidelli, 2003).

 The estimates of the number of objects of the Kuiper belt obtained by different 
authors are given in Table 1.
To obtain estimates of number of Kuiper belt objects (Table~1) and their masses
(Table~2), the authors make some evaluations of the geometric albedo, the parameters
of the luminosity functions, and of the distributions of the KBO by size.
The algorithms used by different authors have differences which can be seen
in the corresponding papers.
 
It is seen from Table 1 that the number of large objects in the Kuiper belt
is about or does not surpass $\sim 10^5 $. Note that in the work (Pitjeva 2010) using dynamic 
estimates, it was shown that the number of objects with D = 100 km and a density 
of 2 $g/cm^3$ should also not exceed $10^5$.

\begin{table}[h]
\caption{Estimates of the number of trans-neptunian objects}
\begin{tabular}{l|c|c|c}
\noalign{\smallskip}
\hline
\noalign{\smallskip}
Year & Author &  Estimate & Note \\
\noalign{\smallskip}
\hline
\noalign{\smallskip}
1998 & Jewitt et al. & $ > 7 \cdot 10^4 $ &  $D > 100$ km    \\
2001b & Trujillo et al. & $\sim  4.7 \cdot 10^4$ & classic KBO, $D > 100$ km \\
2002 & Brunini A. &  $\sim 10^5$ & between 30 au and 50 au, $D > 100$ km\\
2003 & Morbidelli et al. & $\sim 10^5 $ & between 30 au and 55 au, $D > 50$ km   \\
2010 & Pitjeva E. V. &  $< 10^5$  & $\mathrm{D}=100$ km, $\rho = 2$ g/cm$^3$ \\
\noalign{\smallskip}
\hline
\noalign{\smallskip}
\end{tabular}
\end{table}

The first estimates of the mass of the Kuiper belt (Table 2) were obtained statistically. 
Statistical estimates vary widely and are based on different, not entirely 
reliable assumptions about the albedo and density of belt objects. Since there 
are very large uncertainties, the authors publish estimates of the masses without 
their uncertainty. It is seen from Table 2 that the mass of the Kuiper belt is 
within 0.01 -- 0.2 $m_{\oplus}$.

\begin{table}[h]
\caption{Statistical estimates of the mass of the Kuiper Belt}
\begin{tabular}{l|c|c|c}
\noalign{\smallskip}
\hline
\noalign{\smallskip}
Year & Author & Mass & Note \\
\noalign{\smallskip}
\hline
\noalign{\smallskip}
1997 & Weissman, Levison & 0.1 $\div$ 0.3 $m_{\oplus}$ & between 30 au and 50 au \\
1998 & Jewitt et al. & $\sim 0.1$ $m_{\oplus}$ &  \\
1999 & Chiang, Brown & $\sim 0.2$ $m_{\oplus}$ & between 30 au and 50 au \\
1999 & Kenyon, Luu & $\sim 0.1$ $m_{\oplus}$ & between 30 au and 50 au \\
2001 & Gladman et al. & 0.04 $\div$ 0.1 $m_{\oplus}$ & between 30 au and 50 au \\
2002 & Luu, Jewitt & 0.01 $\div$ 0.1 $m_{\oplus}$ & between 35 au and 150 au \\
2002 & Kenyon & 0.1 $\div$ 0.2 $m_{\oplus}$ & total mass beyond the orbit Neptune \\  
2004 & Bernstein et al. & 0.010 $m_{\oplus}$ & classic Kuiper belt \\
2009 & Booth et al. & 0.01 $m_{\oplus}$ & classic objects \\
     &                    & 0.02 $m_{\oplus}$ & scattered objects \\
2010 & Vitense et al. & 0.05 $m_{\oplus}$ & classic + resonant objects \\
     &                    & 0.07 $m_{\oplus}$ & scattered objects \\
 2014&  Fraser et al. &  0.01 $m_{\oplus}$ &  hot populations KBOs\\
 2014 &  ---  &  0.0003 $m_{\oplus}$ &  cold populaions KBOs\\  
\noalign{\smallskip}
\hline
\noalign{\smallskip}
\end{tabular}
\end{table}

The proposed models for the formation of the Kuiper belt require a significant 
initial mass of the belt, of the order of 10--30 $m_{\oplus}$ (Stern 1997; Delsanti 
and Jewitt 2006; Kenyon 2002). Statistical mass estimates (Table 2) at present 
 give a value of 2-3 orders of magnitude less. Therefore, various processes are 
proposed for rapidly dispersing the primordial cloud.

\section{\bf Dynamic estimates of the mass of the Kuiper Belt}

 The population of the Kuiper belt occupies a large volume of outer space, which 
is very sparse. Nevertheless, the presence of the belt, which includes not only 
hundreds of thousands of large objects, but millions of bodies of smaller dimensions, 
leads to a noticeable gravitational effect on the motion of bodies in the Solar 
system, which must be correctly taken into account for modern accuracy of 
observations.

In order to obtain a dynamic estimate of the mass of the Kuiper belt, it was necessary to 
build a dynamical model of the Kuiper belt, include it together with planets in the total 
model for constructing EPM ephemerides, fit the created ephemerides to observations, 
and obtain a dynamic estimate of the Kuiper belt mass from observations.

\subsection{\bf Masses of large bodies of Kuiper belt objects with individual ratings}

The Pluto-Charon system has always been regarded by us as a planet while constructing 
EPM ephemerides.
The equations of motion of the other 30 largest TNOs were also directly included in the 
simultaneous integration with planets and other objects. 
Among these 31 large TNOs, there are 11 objects having satellites.
In this case, their masses are fairly well known from the motion of satellites.
Table 3 gives estimates of the masses of these bodies  
and, therefore, the most accurate masses (taken from astronomical 
data sites). The estimates of the masses in this table are given with an uncertainty 
equal to $ 1 \sigma $, where  $ \sigma $ is the standard error of the least-squares method.

\begin{table}[h]
\begin{tabular}{l|c|l|c}
\hline\noalign{\smallskip}
Object TNO& Mass, $m_{\oplus}$ &  Autor, reference \\
\noalign{\smallskip}
\hline\noalign{\smallskip}
 Eris-136199 & $(27.96 \pm 0.33)\cdot10^{-4}$ &  Brown, Schaller, 2007 \\ 
 Pluto+Charon &  $(24.47 \pm 0.11)\cdot10^{-4}$ &  Brozovic, 2015 \\ 
 Haumea-136108 & $(6.708 \pm 0.067)\cdot10^{-4}$ &  Ragozzine, Brown, 2009 \\
 Makemake-136472 & $(4.35 \pm 0.84)\cdot10^{-4}$ &  Website A \\
 2007 OR$_10$-225088 & $(6.3 \pm 4.5)\cdot10^{-4}$ &  Website B \\
 Quaoar-50000 &  $(1.67 \pm 0.17)\cdot10^{-4}$ &  Fraser et al.,2013 \\ 
 Orcus-90482 & $(1.0589 \pm 0.0084)\cdot10^{-4}$ &  Brown et al.,2010 \\ 
 2003 $AZ_{84}$-208996 & $(0.69 \pm 0.33)\cdot10^{-4}$ &  Website C \\
Salacia-120347 & $(0.733 \pm 0.027)\cdot10^{-4}$ & f Stansberry et al., 2012 \\
Varda-174567 & $(0.446 \pm 0.011)\cdot10^{-4}$ &  Grundy et al., 2015 \\
2002 UX$_{25}$-55637 & $(0.2093 \pm 0.0050)\cdot10^{-4}$ &  Brawn, 2013 \\

 \noalign{\smallskip}\hline
\end{tabular}
\caption{Estimates of the masses of system for the large TNOs with satellites}
\end{table}

  For the remaining 20 objects (out of 31) the masses were calculated from their diameters and densities, these masses estimates are at much worse. The error for each such object, as well as their total mass error, is calculated with taking into account the error in the diameter, as well as the error in the density. Usually, different papers (up to 10) have been studied for the selection of the most reliable values. The average diameter was taken; the diameter uncertainty was taken maximal from available papers. If there were no density estimates, the density was assigned 2.0 g/cm$^3$ with the uncertainty of 1 g/cm$^3$.
 
  The total mass of 31 individually valued large 
objects, including the Pluto / Charon system, is

 $ m_{31TNO} = 0.0086 \pm 0.0017  \ m_{\oplus}  \ (3 \sigma). $

The description of dynamic models for taking into account the gravitational influence
of smaller TNOs is given in the next section, as well as the estimation of their masses.

\subsection{\bf Modeling by a one-dimensional and two-dimensional ring}

 The first dynamic estimates of the mass of TNOs in the Kuiper belt region were 
made in the paper of 2010 (Pitjeva 2010) based on ephemerides EPM2008. 
In that work, the gravitational effect of the aggregate of small or not yet 
discovered bodies of the Kuiper belt was modeled by a one-dimensional material ring with 
a radius of 43 au in the plane of the ecliptic. Later Kuiper belt mass 
estimates were made in a similar manner using ephemerides EPM2011 (Pitjeva 2013) 
and ephemeris EPM2013 (Pitjeva and Pitjev 2014). 
The results for EPM2008--EPM2013 are given in Table 4.  
The total mass is the sum of the masses of the 31 largest TNOs and the estimated mass 
of the stimulated Kuiper ring.

\begin{table}[h]
\caption{Dynamic estimates of the mass of the Kuiper Belt}
\begin{tabular}{l|c|c|c}
\noalign{\smallskip}
\hline
\noalign{\smallskip}
Ephemeris & Mass of  & Total mass of & {\bf Author, reference}\\
          & Kuiper ring  &  Kuiper belt   &       \\
\noalign{\smallskip}
\hline
\noalign{\smallskip}
EPM2008 & $1.66\cdot10^{-2}\ m_{\oplus}$ & $2.58\cdot10^{-2}\ m_{\oplus}$ & Pitjeva 2010 \\ 
EPM2011 & $(1.67\pm 0.83)\cdot10^{-2}\ m_{\oplus}$ & $ 2.63\cdot10^{-2}\ m_{\oplus}$ & Pitjeva, 2013\\
EPM2013 & $(1.08 \pm 0.59)\cdot10^{-2}\ m_{\oplus}$ & $ 1.97\cdot10^{-2}\ m_{\oplus}$ & Pitjeva, Pitjev, 2014\\
EPM2016 &$(1.45 \pm 0.41)\cdot10^{-2}\ m_{\oplus}$ &$(2.28\pm 0.46)\cdot10^{-2}\ m_{\oplus}$ & Pitjeva et al., 2017\\
\noalign{\smallskip}
\hline
\noalign{\smallskip}
\end{tabular}           	
\end{table}

 For a more accurate estimation of the total mass, a transition was made from 
the modeling of the general gravitational attraction of the smaller TNO by a one-dimensional ring
(Ephemerides EPM2008-2015) to the modeling by a two-dimensional homogeneous ring 
with dimensions corresponding to the observed width of the main part of the belt. 
This is a more adequate representation because the most dense part of the 
Kuiper belt lays in an annular zone with a width of more than 8 au between the two main 
orbital resonances with Neptune 3:2 and 2:1 with the corresponding mean distances 
from the Sun $ \sim $ 39.4 au and $ \sim $ 47.8 au. It contains the bulk of the 
population of the Kuiper belt and includes classical objects and the largest number 
of resonant belt objects. 
A more accurate modeling of this component of the belt is essential 
for Neptune, Uranus and Saturn. Neptune's orbit passes relatively close to the inner 
boundary of the belt; and for Saturn, there are very accurate observational data of 
the Cassini spacecraft, which can be used to refine the gravitational influence of 
the belt. 
Since this region represents the most dense part of the belt and beyond 
its limits the number of objects falls significantly, then the distances are 39.4 au
and 47.8 au were taken in our work as, respectively, the internal and external 
boundaries in our modeling by a two-dimensional ring and obtaining an estimate 
of the mass of the main part of the classical and resonant objects of the Kuiper 
belt.

The finding of the gravitational potential and its derivatives for a plane 
two-dimensional homogeneous ring leads to expressions involving complete elliptic 
integrals of the 1st, 2nd, and 3rd kinds, the calculation of which was replaced 
by finding the values of the hypergeometric function of four arguments (Pitjeva and 
Pitjev 2014). The estimates of the mass of the two-dimensional Kuiper ring and 
the total mass of the Kuiper belt obtained for ephemerides EPM2016 is presented in
the last row of Table 4. 

From the test calculations (Table 5) it has been found that the effect of the 
two-dimensional ring is significantly different from that of the one-dimensional ring
of same mass, especially on Uranus and Neptune.

\begin{table}[h]
\caption{Shifts of the planet's perihelion due to the influence of 
the one-dimensional and the two-dimensional ring over 100 years (in arc seconds), 
for the mass ring  $m = 2 \cdot 10^{-2} m_{\oplus}$}
\begin{tabular}{l|c|c|c}
\noalign{\smallskip}
\hline
\noalign{\smallskip}
Planet &One-dimensional&One-dimensional&Two-dimensional\\
\cline{2-4}
 & R=43 au & R=44 au & $R_1=39.4$ au, $R_2=47.8$ au\\
\noalign{\smallskip}
\hline
\noalign{\smallskip}
  Neptune & $0^{\prime\prime}.0437$ & $0^{\prime\prime}.0376$ & $0^{\prime\prime}.0432$\\
  Uranus    & $0^{\prime\prime}.0095$ & $0^{\prime\prime}.0086$ & $0^{\prime\prime}.0091$\\
  Saturn & $0^{\prime\prime}.0024$ & $0^{\prime\prime}.0022$ & $0^{\prime\prime}.0023$\\
  Jupiter & $0^{\prime\prime}.0009$ & $0^{\prime\prime}.0008$ & $0^{\prime\prime}.0009$\\
  Mars   & $0^{\prime\prime}.0003$ & $0^{\prime\prime}.0003$ & $0^{\prime\prime}.0003$\\
\noalign{\smallskip}
\hline
\noalign{\smallskip}
\end{tabular}
\end{table}

Dynamic estimates of the mass of the Kuiper Belt obtained by other authors are 
not yet known.

\subsection{\bf Discrete Rotating Model}

The above mentioned models for the Kuiper belt have a disadvantage when 
taking into account the mutual influence of the planets on the belt, since
the belt should move as a whole. Such movement does not correspond to the actual 
interaction of the planets and the moving bodies of the belt. To avoid this 
drawback,  a new discrete rotating model was proposed with EPM2017.

 The Kuiper belt was modeled by a system of point masses located in the plane 
of the ecliptic,  at the initial moment in circular orbits. 
The point masses do not interact. Gravitational interaction occurs between each of 
these point masses and the Sun, planets and large asteroids. Initial conditions
for the point masses were such that they are located
uniformly on three circular lines. On two 
circles that corresponded to the boundaries of the ``main'' part of the belt
($R_1 = 39.4 $ au and $ R_2 = 48.7 $ au), 40 points were located. The third circle
corresponded to the position of the ``core'' of the belt (Bannister et al. 2016) 
($ R_m = 44 $ au) has 80 point masses, twice the number
of border circles. The velocities of the points corresponded to a circular motion. The total 
mass of the model was a parameter that was determined while fitting ephemerides to the 
observational data of the spacecraft.

\section{\bf Planetary ephemerides EPM2015, EPM2016 and EPM2017}

The results of the dynamical estimation of the mass of the Kuiper Belt in this 
paper were found using a discrete model of moving point masses and are based 
on the processing of observations for the latest version of the ephemerides 
EPM2017 (Ephemerides of Planets and the Moon) of the IAA RAS (Pitjeva and Pavlov 2017). 

EPM ephemerides contain coordinates and velocities of the Sun, the Moon, 
eight major planets, Pluto, three largest asteroids (Ceres, Pallas, Vesta) 
and 4 TNO (Eris, Haumea, Makemake, Sedna), as well as lunar libration and TT-TDB. 
EPM2017 covers the timespan of more than 400 years (1787--2214).
The dynamical model of EPM is based on the Parameterized Post-Newtonian N-body metric 
for General Relativity in the barycentric coordinate system (BCRS) and the TDB time scale.
The motion of the Sun, the planets (including Pluto), and the Moon (as point-masses) obeys 
the Einstein-Infeld-Hoffmann relativistic equations, with additional perturbations from: 
solar oblateness, 301 largest asteroids and 30 largest trans-neptunian objects (TNO), 
as well as two annuli: the first for the asteroid belt, and the second for the Kuiper belt.
The parameters of the lunar and planetary parts of EPM2017 are in agreement with each other.
EPM2017 has been oriented to the ICRF2 with an accuracy better than 0.2 mas ($ 3\sigma $) by 
including into the total solution 266 ICRF2-based VLBI measurements of spacecraft taken 
from 1989-2014 near Venus, Mars, and Saturn.
EPM2017 ephemerides are the most recent in the line of continuous development that
includes EPM2016, EPM2015, and EPM2013 which is described in detail in the paper
(Pitjeva and Pitjev 2014).

The main changes in EPM2017 since EPM2013 are the following:
\begin{itemize}

\item
A revised version of the ERA software package (ERA-8) is used (Pavlov and
Skripnichenko 2014);   

\item
A new model of the orbital-rotational motion of the Moon (Pavlov, Williams 
and Suvorkin 2016)  was implemented, based on the equations used in JPL DE430 
ephemeris (Williams et al. 2013) with combination of up-to-date 
astronomical, geodynamical, and geo- and selenophysical models;

\item
The astronomical unit was recorded in the SI system at a value equal to 
149597870700 ~ m in accordance with Resolution B2 of the XXVIII of the General 
Assembly of IAU (2012), and the gravitational constant of the Sun ($ GM_{Sun}$) 
is determined from observations;

\item
Inclusion of the Lense-Thirring acceleration into model;

\item
Improved relativistic barycenter definition;

\item
Modeling the acceleration of the Sun as a regular body;

\item
The planetary part of EPM2013 was updated by adding two-dimensional rings of 
asteroids (EPM2015) and the Kuiper belt (EPM2016) to the dynamic model of the 
Solar system  (where the maases of these rings are parameters), and for EPM2017 - two discrete rotating rings for the Main belt of 
asteroids and the Kuiper belt and including them in joint integration (see above);

\item
Updated and refined data on a number of asteroids and TNOs;

\item
The number of observations increased: about 800,000 positional observations 
of planets and spacecraft (1913--2015), as well as LLR normal points (1970-2016)
were used, including new infrared LLR data  (Viswanthan et al. 2016),
(Courde et al. 2017) and ranging data from the Cassini and MESSENGE
(Hees et al. 2014)
and the four-year MESSENGER (2011--2015)  (Park et al. 2017) 
spacecraft;

\item
the long Holocene version of ephemeris, EPM2017H.
\end{itemize}
 EPM2017H ephemeris are the longer version of EPM2017, 
covering the timespan of more than 13100
 years (10107 BC -- AD 3036, 
H stands for Holocene) (Cionco and Pavlov 2018). Within the EPM2017 timespan, EPM2017 and EPM2017H 
are practically equal with the only exception of the Moon.
The dynamical model used during the production of EPM2017H differs in two pieces. 
First EPM2017H uses the model of the precession of the Earth valid over thousands of years 
(Vondrak, Capitaine and Wallace 2011). 
Second, the lunar model of EPM2017H does not have friction between crust and core, 
to avoid exponential growing of the angular velocity of the lunar core in the past. 
That decision is similar to what has been done for the DE431 ephemeris model as compared to DE430. 
The lunar parameters of the solution were then refitted to the EPM2017H model.

\subsection{\bf Lense-Thirring acceleration}

The ephemeris EPM2017 takes into account the relativistic acceleration of 
Lense-Thirring:

$$ {{\bf\ddot r}_i}^{LT}={2\over{c^2}}GS_{Sun}{1\over{{r_{iS}^3}}}R_{Sun}
\left({\bf\dot r}_{iS} \times {\bf z} + 3{{{\bf z} \cdot {\bf r}_{iS}}\over
{r^2}}{\bf r}_{iS} \times {\bf\dot r}_{iS} \right)  $$

where:
\begin{itemize}
\item ${\bf r}_{i}$ is the barycentric position of the i-th body
\item ${\bf r}_{iS}$ is the heliocentric position of the i-th body
\item $r_{ij}=|{\bf r}_{j}-{\bf r}_{i}| $
\item {\bf z}=$(0,0,1)$
\item $ R_{Sun}$ -- rotation matrix from $\bf z$ to the Sun's pole. The following 
      value is used: $R_{Sun}=R_z(16^o.13)R_x(26^o.13)$
\item $G$ is the gravitational constant
\item $S_{Sun}$ is the angular momentum of the Sun. Following the decision from 
Park et al. (2017), we take $S_{Sun}=190\times10^{39}$ kg m$^2$/s.
\end{itemize}  

\subsection{\bf Relativistic barycenter definition}

The relativistic barycenter of point-masses $\mu_i$, where $\mu_i= GM_i$ is the 
i-th body's gravitational parameter, is defined as
 
   $$  \bf b = {{\sum_i {\mu_i}^*{\bf r}_{i}}\over{\sum_i {\mu_i}^*}}, $$
where ${\bf r}_{i}$ is the position of the i-th body, while ${\mu_i}^*$ is the 
body's relativistic gravitational parameter:

 $${\mu_i}^* = \mu_i \left(1 + {1\over{2c^2}}{\dot r_i}^2 - {1\over{2c^2}}
\sum_{j \ne i}{{\mu_j}\over{r_{ij}}} \right)$$
Initial positions and velocities of the bodies are equally adjusted so that the 
position of center mass and the momentum are zero:

   $$ \sum_i {\mu_i}^*{\bf r}_{i} = {\bf 0}         $$
   $$ \sum_i \left({\dot \mu_i}^*{\bf r}_{i} + {\mu_i}^*{\bf \dot r}_i\right) = {\bf 0}  $$
In previous versions of EPM, the ${\dot \mu_i}^*$ was neglected.
  $\mathbf{{\mu_i}^*}$ in the barycenter of the Solar system was firstly presented
 in (Fienga et al. 2008).

Control of the position of the barycenter was carried out taking into account 
the simulated rings of the Main Asteroid Belt and the Kuiper Belt. The 
position of the center of gravity of the Solar system, taking into account the 
Sun, planets, large asteroids and TNO, and discrete models for the Main Asteroid belt 
and the Kuiper Belt, remained practically at  the origin of coordinates.
However, numerical round-off errors in the integration cause the barycenter, 
initially placed at the origin, to drift at a slow pace. 
In the timespan of 1900-2020, the displacement of the barycenter from the origin 
is less than 0.03 mm.

\subsection{\bf Modeling the acceleration of the Sun as a regular body}

In earlier versions of the EPM, the Sun was explicitly placed to keep the barycenter at the origin. 
In EPM2017, the Sun obeys the Einstein-Infeld-Hoffmann equations of motion as all the other bodies.
Similar approach has been taken in DE ephemeris starting from DE430 (Folkner et al. 2014)
and INPOP ephemeris starting from the first release INPOP06 (Fienga et al. 2008).
The perturbations from solar oblateness and Lense-Thirring effect act not only on planets, 
but on the Sun, too, following Newton's third law.
The mutual influence of the asteroid belt and the Kuiper belt to the orbit of the Sun, 
neglected in previous versions of EPM, is properly taken into account in EPM2017.

\subsection{\bf Recent masses and orbits of some asteroids}

Some asteroid data (orbits and masses) were updated. In particular, the new mass value was obtained 
during the approach of the Dawn spacecraft to Ceres (Park, et al. 2016). 
The masses of 30 large asteroids (not including 13 double and multiple asteroids) were improved 
from observations of different spacecraft. 
Masses of other 255 asteroids were calculated from their diameters and densities.

\subsection{\bf Observations for EPM2017}

New LLR data (till the end of 2016) was added into EPM2017 solution---most notably, 
the new high quality infrared data from the OCA observatory (Viswanthan et al., 2016), (Courde et al., 2017).

The planetary part of EPM2017 ephemerides has been fitted to about 800000 observations 
of different types, spanning 1913--2015, from classical meridian observations 
to modern planetary and spacecraft ranging.
However, spacecraft ranging near planets are correlated with each other on each tracking pass,
so there is effectively only one independent range point for each pass. 
Based on this, Dr. Folker (JPL) formed normal points of JPL spacecraft ranging; 
the normal points for other spacecraft ranging were formed by authors. 
Moreover, we deleted points where the Mercury-Sun-Earth angle was larger than 120 degrees due to solar plasma giving a large noisy effect, as Dr. Folker did with the MESSENGER data. I It is the filtr for Messenger. For Cassini only the 3-sigma criterion was used, i.e. several points were thrown out if their the value (0-C) exceeded the errors by a factor of 3. 

The new radar data were obtained due to the courtesy of William Folkner (JPL) and 
Agnes Fienga (IMCCE) via private communication and the NASA JPL  
Observational Data for Planets and Planetary Satellites (Folkner 2015), 
and  INPOP Astrometric Planetary Data(Fienga 2017) webpages.

For the development of EPM after EPM2013, data for the planets were added including 
radar observations received in 2010--2014: for Odyssey, Mars Reconnaissance Orbiter (MRO), 
Mars Express (MEX) and Venus Express (VEX), as well as VLBI data (2011--2014) 
for VEX, Odyssey, MRO, Cassini. 
The most valuable for us have been the recent data the ten-year Cassini 2004--2014
(Hees et al. 2014)
and the four-year MESSENGER (2011--2015) (Park et al. 2017). 
In Figure 1 and Figure 2, the one-way residuals of Cassini and Messenger calculated for EPM2017
are shown. 

As a result, the Mercury and Saturn ephemeris has become significantly more accurate than 
it was in EPM2013.  

\begin{figure}[h]
\caption{The residuals of one-way
ranging for spacecraft Cassini computed using EPM2017, the wrms is  20.17 metres}
\includegraphics[scale=1.2]{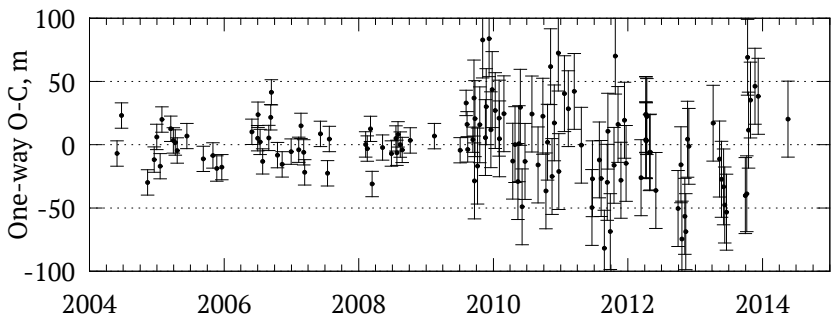}
\end{figure}

\begin{figure}[h]
\caption{The residuals of one-way
ranging for spacecraft MESSENGER computed using EPM2017, the wrms is 0.7 metres}
\includegraphics[scale=1.2]{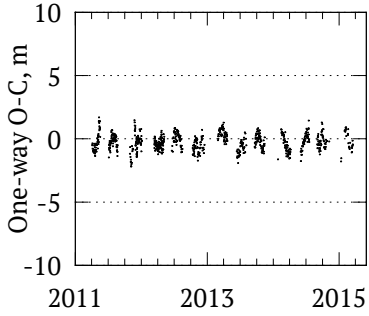}
\end{figure}

Moreover, new CCD observations were added obtained in 2012-2013 at Flagstaff and TMO observatories, 
and new data for Pluto obtained in 1950-2013 at Brazilian Pico dos Dias observatory (Benedetti-Rossi 
et al. 2014), and a new analysis of photographic plates taken at Lowell Observatory from 1931 to 1951 
(Buie and Folkner 2015). 
In total, 53368 normal points of spacecraft ranging and 72929 optical observations 
have been used for fitting ephemerides and estimating the parameters of EPM2017.

The observations ranging of spacecraft located near planets and the Martian landers ---
Viking-1, Viking-2, Pathfinder have been used for estimating the mass of the Kuiper belt.
The number of observations (for the Martian landers) and  normal places, as well as intervals of 
these observations, are shown in Table 6.

\begin{table}[h]
\caption{The ranging of the spacecraft and Martian landers used for estimating the ring masses}
\begin{tabular}{c|c|c|c}
\noalign{\smallskip}
\hline
\noalign{\smallskip}
Planet & Spacecraft, landers & Interval of & Number of observ. \\
      & & observations &  or norm. points \\
\noalign{\smallskip}
\hline
\noalign{\smallskip}
Mercury & MESSENGER & 2011--2015 & 813 \\
Venus   & VEX &       2006--2013 & 2520 \\
Mars    & Viking-1 & 1976--1982 & 1178 \\
        & Viking-2 & 1976--1977	&  80 \\
 &	Pathfinder &	1997 &	90  \\
 &	MGS	 & 1998--2006 &	 7341 \\
 &	Odyssey	 & 2002--2013 &	7323 \\
 &	MRO	& 2006--2013 &	2169 \\
 &      MEX &	 2009--2014 &	3587 \\
Saturn & Cassini & 2004--2014 &	161  \\
\hline
\end{tabular}
\end{table}

The Cassini observations play an important role in the results obtained, although 
all observations of planets and spacecraft were used in the processing. 
The accuracy of modern optical CCD observations 
reaches tens of milliarcseconds. The trajectory observations of spacecraft are much more accurate 
--- about 20 meters at the distance of Saturn and 1--2 meters for Mars. 
For Uranus, optical observations cover an 
interval of slightly more than one revolution with   acceptable accuracy of 0.5" and better
, and for 
Neptune, slightly more than half of the total revolution around the Sun. From 
radar measurements, there is only two 3-D points, one for Uranus and one for Neptune, 
obtained during the flight of the Voyager-2 spacecraft. 

\subsection{\bf The long Holocene version of ephemeris, EPM2017H}

EPM2017H ephemeris are the longer version of EPM2017, 
covering the timespan of more than 13100
 years (10107 BC -- AD 3036, 
H stands for Holocene) (Cionco and Pavlov 2018). Within the EPM2017 timespan, EPM2017 and EPM2017H 
are practically equal with the only exception of the Moon.
The dynamical model used during the production of EPM2017H differs in two pieces. 
First EPM2017H uses the model of the precession of the Earth valid over thousands of years 
(Vondrak, Capitaine and Wallace 2011). 
Second, the lunar model of EPM2017H does not have friction between crust and core, 
to avoid exponential growing of the angular velocity of the lunar core in the past. 
That decision is similar to what has been done for the DE431 ephemeris model as compared to DE430. 
The lunar parameters of the solution were then refitted to the EPM2017H model.

\section{\bf The total mass of the Kuiper Belt}

The total mass of the 31 individually estimated of the Kuiper belt is
$$ M_{31} = (0.86 \pm 0.17) \cdot 10^{-2} \ m_{\oplus}\ \ (3 \sigma). $$

  The estimation of the total mass of the smaller or still undetected 
bodies of the Kuiper belt, as well as their small fragments, has obtained from the modeling fluence by
of their general gravitational influence by the discrete rotating ring ($R_1, R_2$). 

  The mass of the simulated part of the belt is
$$      M_{ring} = (1.108 \pm 0.25) \cdot 10^{-2} \ m_{\oplus}\ \ (3 \sigma).  $$

Thus, the final result for the total mass of the belt including 
large and small bodies, as well as undiscovered objects, is equal 

   $$ M_{belt}= (1.97 \pm 0.35) \cdot 10^{-2} \ m_{\oplus}\qquad (3 \sigma) . $$
 These values of the 31 TNO masses were fixed when obtaining the estimate of the total mass. The mass of the ring was obtained with its own uncertainty. The total mass of the ring includes ring mass uncertainty and 31 TNO mass uncertainty.

To obtain a more reliable estimate of the mass of the modeling ring, in addition to the 
main version of the ring with dimensions 
$R_1 = 39.4$ au, $R_2 = 48.7$ au, the test version was considered when all the belt 
objects,including Pluto, were assigned to the simulated ring, and the large objects were 
not separately considered. In this version the total mass of the Kuiper belt has turned 
the same as in the main version.

Note that, unlike the Main asteroid belt, the total mass of large objects of 
the Kuiper belt including Pluto-Charon is about $ 40 \% $, the rest falls on 
smaller bodies and undiscovered objects.

The error in determining the mass of the Kuiper belt, as well as residuals (O-C)
have decreased with the discrete model.
 The uncertainty in the total mass has decreased due to using the new model of th{e Kuiper belt and the parameters – the mass of modeling ring and the total mass of the Kuiper belt were obtained using the new model. The process of transition from the one-dimensional ring model to the two-dimensional discrete ring model, as well as the deficiencies of previous models, are described in the paper. Each next model has advantages - it more adequately represents the gravitational perturbation from the 
ring to the planets.  

\section{The influence of TNOs on the residuals of Cassini}

 The aim of the work is to improve and refine the dynamic model of the 
Solar system to obtain exact ephemerides of the planets and more adequately 
take into account the influence of a large number of distant bodies concentrated behind the orbit of Neptune.
The Kuiper Belt truly exists. The number of discovered TNOs 
currently exceeds 2000 and continues to grow. There are  bodies with the diameter 100 km or bigger, some of them of objects comparable 
in size and mass to Pluto, as well as there are a lot of smaller objects, 
not yet open, yet uncovered. The total mass according to statistical estimates 
is much higher the total mass of the asteroid belt.

It is necessary to include the gravitational attraction of trans-Neptunian 
objects (TNOs), primarily to the planets Neptune, Uranus, Saturn, the 
closest to Kuiper belt.  
Although Saturn is still far from the Kuiper belt, and  most 
of the deviations due to Kuiper belt perturbations are absorbed by 
the fitting the orbital elements of Saturn, but some improvements 
using perturbations from the Kuiper belt are nevertheless observed.
We constructed the test version EPM with the zero masses all TNO objects, 
including the mass of the modeled TNO ring (EPM2017t), and then improve this version 
to all observations.
In Table 7, there are residuals (their standard deviations) of the Cassini data 
for ephemeris version constructed with TNO(EPM2017), residuals for ephemeris without 
TNO (EPM2017t), and, for comparision, standard deviations of Cassini ranging were 
taken out from the ephemeris INPOP2017 by Viswanathan et al. (2017), constructed 
without TNO. Moreover, there are standard deviations of VLBI Cassini obtained from
the same ephemerides for right ascension.    

\begin{table}[h]
\caption{The influence of TNOs on the residues of Cassini}
\begin{tabular}{l|c|c}
\noalign{\smallskip}
\hline
\noalign{\smallskip}
Ephemerides) & Cassini ranging residuals & Cassini VLBI residuals\\
\noalign{\smallskip}
\hline
\noalign{\smallskip}
 EPM2017 (with TNO's) & 20.170 m & 0.183 mas \\
 EPM2017T (without TNO's) & 22.734 m &  0.366 mas \\
 INPOP17a (without TNO's) & 31.613 m & 0.553 mas \\
  \noalign{\smallskip}
\hline
\noalign{\smallskip}
\end{tabular}
\end{table}

Perturbations of TNO affect mostly the motion of the perihelon of the planets, i.e.
right ascension, so all three values for standard deviations for declination 
are approximately the same.
The inclusion of attraction from the Kuiper belt to the dynamic model of the 
Solar system is especially important for constructing exact ephemerides
of Saturn and future exact ephemerides for Uranus and Neptune. 

\section{Gravitational Influences of the Kuiper Belt and the 9th Planet}

It was found that the total mass of the Kuiper belt is two orders of magnitude larger than 
the mass of the Main asteroid belt (Pitjeva and Pitjev 2014; Pitjeva et al. 2017), and at 
the same time is almost three orders of magnitude smaller than the mass of the 
proposed 9th planet (Batygin and Brown 2016). According to the peculiarities of 
the distribution of the orbital elements of some numbers of comparatively large 
objects of the Kuiper belt, they proposed the existence of a large trans-neptunian 
planet with a mass of $\sim 10 \, m_{\oplus}$ and having an elongated, inclined 
orbit ($ a = 700 $ au, $ E = 0.6 $, $ i = 30^o $). In the works (Fienga et al. 
2016; Folkner et al. 2016; Millholland et al. 2017), the possible influence of 
a hypothetical planet on the motion of Saturn from high-precision observations
of the Cassini spacecraft, was investigated. Those authors claim that if the ninth planet 
with such a mass exists, it is currently in the aphelion of the orbit 
($\sim$ 800--1100 au).

It should be noted that the total gravitational acceleration for Saturn created 
by the Kuiper belt is also directed toward the periphery and can be comparable 
or even exceed the perturbation from a hypothetical remote planet with a mass of 
the order of the 10 Earth masses.
Comparison of the perturbing accelerations on the planets from the Kuiper belt, which has the mass found in this paper (minus the mass of Pluto taken into account in ephemerides) m = 0.0175 $m_{\oplus}$ and the acceleration from the hypothetical planet with a mass of 10 $m_{\oplus}$ was made. It was found that the distance at which these accelerations coincide at the maximum acceleration from the 9th planet, that is, when the planet and 9th planets are on one side of the Sun on one straight line. At a farther position of the 9th planet, which is assumed by most authors, the acceleration from the Kuiper belt exceeds the acceleration caused by the 9th planet.

Since in the aforementioned works, the total gravitational attraction of the
Kuiper belt was not taken into account, we compared the perturbations exerted by the 
hypothetical ninth planet and the Kuiper belt. As parameters of the belt modeled 
by the discrete moving point masses (subsection 3.3), the values $ m = 0.02 
\, m_{\oplus} $, $R_1 = 39.4$ au and $R_2 = 48.7$ au were used. 
Various values of the 9th planet's distance from the Sun,  $R$, were considered
because the exact position of a hypothetical planet is not known.
The results are shown in Table 8, where the distances $R$ are given for the 
hypothetical planet, when the magnitude of the acceleration to the planet from 
the Kuiper belt and the perturbing acceleration from the 9th planet becomes 
equal. However, the TNO object -- the Pluto-Charon system is was included into  
equations of motion for integrating together with the planets in the such 
ephemerides as DE and INPOP. So in this case, the neglected mass of the Kuiper 
belt becomes less (the 3-rd column of Table 8), and the distance where the equality 
of the perturbing acceleration from the hypothetical planet and the acceleration 
from the Kuiper belt becomes somewhat large.

\begin{table}[h]
\begin{tabular}{l|c|c}
\noalign{\smallskip}
\hline
\noalign{\smallskip}
Planet &  R (au) & R (au) \\
       &  $m_{belt}=0.02 \ m_{\oplus}$ & $m_{belt}=0.0175 \ m_{\oplus}$\\
\noalign{\smallskip}
\hline
\noalign{\smallskip}
  Neptune &   440  & 462 \\
  Uranus &  515 & 537\\
  Saturn &  540 & 568\\
  Jupiter &  545 & 575\\
  Mars  &  550  & 582\\
\noalign{\smallskip}
\hline
\noalign{\smallskip}
\end{tabular}
\caption{The distances $R$, where the equality of the perturbing acceleration from 
the hypothetical planet and the acceleration from the Kuiper belt, the 9-th planet mass=$10 \ m_{\oplus}$}
\end{table}

With the obtained estimate of the total mass of the Kuiper belt, the gravitational 
perturbation of the planets from it is comparable to the perturbation from the 
hypothetical 9th large planet at distances of the order of 550 au.  
At distances greater than 800 au estimating for its position in the 
aphelion region, the influence from the Kuiper belt to planets exceeds the influence 
from the 9th planet by several times. 
Therefore, when processing observations, it is necessary, at first, take into account 
the gravitational attraction from the Kuiper belt, and then investigate residual discrepancies 
for assessing the possible impact of the distant large planet.

\section{Conclusion}

 The mass of the Kuiper Belt in the Solar System has been estimated on the basis of 
the new version of the ephemerides of planets and the Moon --- EPM2017 using about 800,000 positional 
observations of planets and spacecraft (1913-2015). To more accurately estimate 
the total mass of the Kuiper belt, the transition was made to the representation 
of its total gravitational influence by the a discrete rotating model with radial dimensions 
corresponding to the width of the distribution of classical and resonant belt 
objects (39.4, 47.8 au). This leads to the better representation of the observations 
and decreasing the error in determining the total mass of the belt. The found 
value of the total mass of the Kuiper belt is

$$M_{belt} = (1.97 \pm 0.35)\cdot 10^{-2} \ m_{\oplus}.$$

A correct account of the total gravitational influence of numerous bodies of the 
Kuiper belt is necessary both for constructing a more accurate dynamic model of 
the solar system and high-precision ephemerides of planets.  It is also useful
for a more reliable search for a distant hypothetical planet.

The total perturbation from the Kuiper belt to the planets is comparable in 
magnitude with of the total  gravitational influence of a hypothetical planet with 
gravitational influence of a hypothetical planet with a mass of 10 $ m_{\oplus} $
at distances more than 550 au. 
At distances greater than 800 au corresponding to 
the position of this planet in the aphelion region, the attraction of the belt 
exceeds its perturbation to planets by several times. Therefore, it is necessary 
to take into account, as 
accurately as possible, the gravitational influence of the Kuiper belt when 
processing observations, and only then to investigate residual discrepancies to 
discover the possible influence of the distant large planet.
 
\begin{acknowledgements}
The authors are very grateful to Dr. Dmitry Pavlov for the software for the development of 
the EPM2017 ephemerides (inclusion of the Lense-Thirring acceleration into model,
improved relativistic barycenter definition, modeling the acceleration of the Sun as a regular body,
integration of isochronous derivatives). 
They would like also to thank Mariya Bodunova for calculation of the influence of rings and 
the 9th planet on other planets.
\end{acknowledgements}

The authors declare that they have no conflict of interest.


\begin{thebibliography}{}

\bibitem{}
Allen, R., Bernstein, G., Malhotra, R.: Observational limits on a distant Cold 
Kuiper belt. Astronomical Journal 124, 2949-2954 (2002)

\bibitem{}
Bannister, M., Kavelaars. J., Petit, J.-M., et al.: The outer solar system 
origins survey. I. Design and first-quarter discoveries. Astronomical Journal
152(3), article id. 70, 25 pp. (2016)

\bibitem{}
Batygin, K., Brown, M.: Evidence for a distant giant planet in the solar 
system. Astron. J. 151(2), article id. 22, 12 pp. (2016) 

\bibitem{}
Benedetti-Rossi, G., Vieira, M. R., Camargo, J. I. B., et al.: Pluto: improved 
astrometry from 19 years of observations. Astron. Astrophys. 570, A86 (2014)

\bibitem{}
Bernstein, G. M., Trilling, D. E., Allen, R. L., et al.: The size distribution 
of trans-neptunian bodies. 
Astron. J. 128(3) 1364-1390 (2004) 

\bibitem{}
Booth, M., Wyatt, M. C., Morbidelli, A., et al.: The history of the solar system's debris disc: observable 
properties of the Kuiper belt. Mon. Not. R. Astron. Soc. 399, 385-398 (2009)

\bibitem{}
Brozovic M., Showalter M. R., Robert A. Jacobson R. A.,  Buie M. W.
The orbits and masses of satellites of Pluto. Icarus 246,  317-329
(2015)

\bibitem{}                                             
Brown, M. E., Schaller, E. L. The Mass of Dwarf Planet Eris. Science. 316 (5831), 1585. 
(2007)

\bibitem{}      
Brown, M. E.,; Ragozzine, D., Stansberry, J., Fraser, W. C. The size, density, 
and formation of the Orcus-Vanth system in the Kuiper belt. The Astron. J. 139, 2700-2705 (2010)

\bibitem{}
Brown M.E. The density of mid-sized Kuiper belt object 2002 UX25 
and the formation of the dwarf planets". Astrophys. J. Letters. 778 (2), L34 
(2013)

\bibitem{}
Brunini, A.: Dynamics of the Edgeworth-Kuiper belt beyond 50 au. Spread of 
a primordial thin disk. Astron. Astrophys. 394, 1129-1134 (2002)

\bibitem{}
Buie, M. W., Folkner, W. M.: Astrometry of Pluto from 1930-1951 observations: 
the Lampland plate collection. Astron. J. 149(1), article id. 22, 13 pp. (2015)

\bibitem{}
Chiang, E., Brown, M. E.: KECK pencil-beam survey for faint Kuiper belt objects.
Astron. J. 118, 1411-1422 (1999)

\bibitem{}
Cionco, R. G., Pavlov, D. A.: The Solar barycentric dynamics from a new solar-planetary ephemeris. 
accepted to Publication in Astron. Astrophys. (2018) 

Lunar laser ranging in infrared at the Grasse laser station
Authors:	
Courde, C., Torre, J. M., Samain, E., Martinot-Lagarde, G., Aimar, M.,
Albanese, D., Exertier, P., Fienga, A., et al.
Lunar laser ranging in infrared at the Grasse laser station
Astron. Astrophys., Vol. 602, id.A90, 12 pp. (2017)

\bibitem{}
De Sanctis, M., Capria, M., Coradini, A.: Thermal evolution and differentiation 
of Edgeworth-Kuiper belt objects. Astronomical Journal 121(5), 2792-2799 (2001)

\bibitem{}
Delsanti, A., Jewitt, D.: The solar system beyond the planets, p.267. In book 
"Solar System Update", P. Blondel and J. Mason. (ed.), Springer, Berlin (2006)
 
\bibitem{}
Elliot, J., Kern, S., Clancy, K., et al.: The Deep ecliptic survey: a search for 
Kuiper belt objects and centaurs. II. Dynamical classification, the Kuiper belt 
plane, and the core population. Astron. J. 129, 1117-1162 (2005) 

\bibitem{}
Fienga, A., Manche, H., Laskar, J., Gastineau, M.: INPOP06: a new numerical planetary 
ephemeris. Astron. Astrophys. 477(1), 315-327 (2008)

\bibitem{}
Fienga, A., Laskar, J., Manche, H., Gastineau, M.: Constraints on the location of 
a possible 9th planet provided by the Cassini data. Astron. Astrophys. 587, id.L8, 4 pp. (2016)

\bibitem{}
Fienga, A.: INPOP Astrometric Planetary Data.
 http://www.geoazur.fr/astrogeo/?href= observations/base (2017)

\bibitem{}
Folkner, W. M., James G. Williams J. G., Boggs D. H., et al.: The Planetary and Lunar 
Ephemeris DE430 and DE431, JPL IPN Progress Report, 42-196 (2014)

\bibitem{}
Folkner, W. M.: Observational data for planets and planetary satellites.
https://ssd.jpl.nasa.gov/?eph-data (2015)

\bibitem{}
Folkner, W., Jacobson, R., Park, R., Williams, J.: Sensitivity of Saturn's Orbit 
to a Hypothetical Distant Planet. AAS, DPS meeting 48, id. 120.07 (2016).

\bibitem{}
Fraser, W. C., Batygin, K., Brown, M. E., Bouchez, A. 
The mass, orbit, and tidal evolution of the Quaoar-Weywot system. 
Icarus. 222 (1), 357-363 (2013) 

\bibitem{}
Fraser, W.C., Brown, M.E., Morbidelli, A., Parker, A., Batygin, K.:
The absolute magnitude distribution of Kuiper belt objects. Astroph. J. 782(2), 100 (2014).


\bibitem{}
Gladman, B., Kavelaars, J., Petit, J.-M., et al.: The structure of the Kuiper belt:
size distribution and radial extent.
Astron. J. 122, 1051-1066 (2001)

\bibitem{}
Gladman, B. Nomenclature in Kuiper belt. Highlights of Astronomy 12, 
193-198 (2002)

\bibitem{}
Grundy, W. M., Porter, S. B., Benecchi, S. D., Roe, H. G., Noll, K. S., 
Trujillo, C. A., et al. The mutual orbit, mass, and density of the large 
transneptunian binary system Varda and Ilmare. Icarus. 257, 130-138 
(2015)

\bibitem{}
Hees, A., Folkner, W., Jacobson, R., Park, R.: Constraints on modified Newtonian
dynamics theories from radio tracking data of the Cassini spacecraft. Pys. Rev. D 
89(10), id.102002  (2014)

\bibitem{}
Jewitt, D., Luu, J. C., Trujillo, C.: Large Kuiper belt objects: The Mauna Kea 
8K CCD Survey. Astronomical Journal 115, 2125-2135 (1998)

\bibitem{}
Kenyon, S.: Planet formation in the outer solar system. Public.
Astron. Soc. Pacific, 114: 265-283 (2002)

\bibitem{}
Kenyon, S. J., Luu, J.: Accretion in the early Kuiper belt. II. Fragmentation.  
Astron. J. 118(2), 1101-1119 (1999) 

\bibitem{}
Levison, H., Morbidelli, A.: The formation of the Kuiper belt by the outward 
transport of bodies during Neptune's migration. Nature,  426(6965), 419-421 (2003)

\bibitem{}
Luu, J., Jewitt, D.: Kuiper belt objects: relics from the accretion disk of the Sun.
Annual Review Astron. Astrophys. 40, 63-101 (2002)

\bibitem{}
Millholland, S., Laughlin, G.: Constraints on planet nine's orbit and sky 
position within a framework of mean motion resonances. Astronomical Journal 153(3),
article 91, 13 pp. (2017)

\bibitem{}
Morbidelli, A., Brown, M, Levison, H.: The Kuiper belt and its primordial 
sculpting.  Earth, Moon, and Planets  92(1), 1-27 (2003)

\bibitem{}
Park, R. S., Konopliv, A. S., Bills, B. G., et al.: A partially differentiated interior 
for (1) Ceres deduced from its gravity field and shape. Nature (Letter), Vol. 537, 
515-522 (2016), doi:10.1038/nature18955

\bibitem{}
Park, R., Folkner, W., Konopliv, et al.: Precession of Mercury's perihelion from 
ranging to the MESSENGER Spacecraft. Astron. J. 153, 121, 7 pp.  (2017)

\bibitem{}
Pavlov. D., Skripnichenko, V.: Rework of the ERA software system: ERA-8 // 
Proc. Conf. Journees - 2014 Systemes de rifference spatio-temporels / eds.: 
Z. Malkin, N. Capitaine.  St. Petersburg,  243-246 (2015)

\bibitem{}
Pavlov, D., Williams, J., Suvorkin, V.: Determining parameters of Moon's orbital 
and rotational motion from LLR observations using GRAIL and IERS-recommended 
models. Celest. Mech. Dyn. Astron. 126(1-3), 61-88 (2016)

\bibitem{}
Petit, J.-M., Kavelaars, J.J., Gladman, B.J. et al.:
The Canada-France ecliptic plane survey-full data release: the orbital structure 
of the Kuiper belt. Astron. J. 142(4), article id. 131, 24 pp. (2011)

\bibitem{}
Pitjeva, E.: Influence of trans-neptunian objects on motion of major
planets and limitation on the total TNO mass from planet and spacecraft. 
Proc. IAU Symp. No. 263 / Icy bodies of the solar system.
D. Lazzaro, D. Prialnik, R. Schulz, J.A. Fernandez (eds.),
Cambridge University Press,  93-97 (2010)

\bibitem{}
Pitjeva, E. V.: Updated IAA RAS planetary ephemerides -- EPM2011 and their use in 
scientific research. Solar System Research 47(5), 386-402 (2013)

\bibitem{}
Pitjeva, E. V., Pitjev N. P.: Development of planetary ephemerides EPM and their 
applications. Celest. Mech. Dyn. Astron. 119, 237-256 (2014)

\bibitem{}
Pitjeva, E. V., Pitjev, N. P., Pavlov, D. A., Bodunova, M. A.: Two-dimensional Annuli 
of the main asteroid belt and trans-neptunian Objects and their influence on the motion of planets.
Trudy Inst. Appl. Astron. Russ. Acad. Sci. 42, 8 pp. (in print) (2017)

\bibitem{}
Pitjeva, E. V., Pavlov, D. A.: EPM2017 and EPM2017H.
 http://iaaras.ru/en/dept/ephemeris/epm/2017/, Accessed 7 November 2017

\bibitem{}
Ragozzine, D., Brown, M. E. Orbits and Masses of the Satellites of 
the Dwarf Planet Haumea = 2003 EL61.  Astron. J. 137 (6), 4766-4776 
(2009)

\bibitem{}
Stern, S., Colwell, J.: Collisional erosion in the primordial Edgeworth-Kuiper 
belt and the generation of the 30-50 au Kuiper gap. Astrophys. J. 490(2): 879-882 (1997) 

\bibitem{}
Stansberry, J. A., Grundy, W. M., Mueller, M., et al. Physical Properties 
of Trans-Neptunian Binaries (120347) Salacia, Actaea and (42355)  Typhon-Echidna.
Icarus. 219, 676-688 (2012)
                            
\bibitem{}
Trujillo, C. A., Brown, M. E.: The radial distribution of the Kuiper belt. 
Astrophys. J. 554, L95-L98 (2001a)

\bibitem{}
Trujillo, C., Luu, J., Bosh, A., Elliot, J.: Large bodies in the Kuiper Belt. 
Astron. J. 122, 2740-2748 (2001b)

\bibitem{}
Viswanthan, V., Fienga, A., Manche, H., et al.: New results for the INPOP lunar ephemerides: 
new modelings for the inner structure and IR LLR data.
The 2016 Internatio
nal Workshop on Laser Ranging, Potsdam, Germany, October 9-14 (2016)

\bibitem{}
Viswanathan,v. Fienga, A., Gastineau, M.,  Laskar, J.INPOP17a planetary ephemerides
Technical Report, 
https://www.researchgate.net/publication/320035644\_INPOP17a\_planetar\_ephemerides
(2017)

bibitem{}
https://cddis.nasa.gov/lw20/docs/2016/papers/31-Fienga\_paper.pdf
\bibitem{}
Vitense, C., Krivov, A., Lohne, T.: The Edgeworth-Kuiper debris disk. 
Astron. Astrophys. 520(A32), 18 pp. (2010)

\bibitem{}
Vondrak, J., Capitaine, N., Wallace, P.: New precession expressions, valid for long time intervals.
Astron. Astrophys. 534, A22 (2011)

\bibitem{}
Williams, J. G., Boggs, D. H., Folkner, W. M.: DE430 Lunar orbit, physical libration, 
and surface coordinates. Jet Propulsion Laboratory Interoffice Memorandum 335-JW, DB, 
WF-20130722-016, California Institute of Technology (2013)

\bibitem{}
Website A http://lnfm1.sai.msu.ru/neb/rw/natsat/double/Makemake.htm

\bibitem{}
Website B https://de.wikipedia.org/wiki/(225088)\_2007\_OR10

\bibitem{}
Website C https://es.wikipedia.org/wiki/(208996)\_2003\_AZ84


\bibitem{}
Weissman, P., Levison, H.: The population of the trans-neptunian region: the 
Pluto-Charon environment. In:  Pluto and Charon, S. Alan Stern,  
David J. Tholen (eds.). Tucson, University of Arizona Press (1997)

\end{thebibliography}
\end{document}